\documentclass[useAMS,usenatbib]{mn2e}
\usepackage{graphics}
\usepackage{epsfig}
\title[X-ray spectral evolution of LMC X-2]{X-ray spectral evolution of the extragalactic Z-source, LMC X-2}
\author[V. K. Agrawal and R. Misra]{V. K. Agrawal$^{1,2}$\thanks{E-mail:
vivek@tifr.res.in} and R. Misra$^3$ \\
$^{1}$ ISRO Headquarters, Bangalore, India \\
$^{2}$ Tata Institute of Fundamental Research, Mumbai, India \\
$^{3}$ Inter University Center for Astronomy and Astrophysics, Pune, India   }
\begin{document}

\date{}

\pagerange{\pageref{firstpage}--\pageref{lastpage}} \pubyear{2009}

\maketitle

\label{firstpage}

\begin{abstract}
We present the results obtained by a detailed study of the extragalactic
Z source, LMC X-2, using broad band Suzaku data and a large ($ \sim 750$ ksec)
data set obtained with the proportional counter array (PCA) onboard RXTE. The
PCA data allows for studying the complete spectral evolution along
the horizontal, normal and flaring branches of the Z-track. Comparison
with previous study show that the details of spectral evolution (like
variation of Comptonizing electron temperature), is similar to that of
GX 17+2 but unlike that of Cyg X-2 and GX 349+2. This suggests that Z
sources are heterogeneous group with perhaps LMC X-2 and GX 17+2 being
member of a subclass. However non monotonic evolution of the Compton
y-parameter seems to be generic to all sources. The broad band {\it
Suzaku} data reveals that the additional soft component of the source
modelled as a disk blackbody emission is strongly preferred over one where
it is taken to be a blackbody spectrum. This component as well as the
temperature of seed photons do not vary when source goes into a flaring
mode and the entire variation can be ascribed to the Comptonizing cloud. The
bolometric unabsorbed luminosity of the source is well constrained to
be  $ \sim 2.23 \times 10^{38}$ ergs/sec which if the source is Eddington
limited implies a neutron star mass of 1.6 M$_\odot$. We discuss the
implications of these results.
\end{abstract}

\begin{keywords}
accretion,accretion discs - X-rays:binaries - X-rays:individual:LMC X-2
\end{keywords}

\section{Introduction}
Low-mass X-ray binaries containing  accreting weakly magnetised neutron
stars exhibit  luminosity and spectral variations on time scale of
 hours to days and are usually divided into two classes:
Z and atoll sources \citep{Has89}. This classification relies upon the path
traced out by individual sources in the X-ray colour-colour diagram (CD)
or hardness-intensity diagram (HID). CD is constructed by plotting the
hard colours (ratio of count rates in two largest energy bands) against
soft colours (ratio of count rates in two lowest energy bands). HID
is constructed by plotting hard colour against intensity. Z-sources
trace out an approximate `Z' shaped path in the CD and HID with the
three branches named as the horizontal branch (HB), the normal branch (NB)
and the flaring branch (FB). Spectral analysis of UV data which is modelled as
reproduced X-ray emission from an outer disk, indicates that the bolometric 
luminosity and hence the mass accretion rate increases
along the Z-track from HB to FB through NB \citep{Has90,Vrt90}.
However, more direct estimation of the bolometric luminosity has proved to
be difficult since that requires both  broad band data and a reliable distance
estimate. Nevertheless, there are indications that Z-sources have
luminosities comparable to the Eddington limit. 
There are nine Z-sources known, including the recent discovery of a
transient galactic X-ray binary \citep{Hom07}. Two of them LMC X-2
\citep{Sma00,Sma03} and RX J0042.6+4115 \citep{Bar03} 
are extragalactic ones.

The radiative processes that produce the X-ray spectra of neutron star
LMXBs are not well constrained. While, there have been attempts to 
have detailed theoretical
model based on the accretion flow \citep[e.g.][]{Psa95}, in general, 
two different 
phenomenological approaches have been adopted to model their spectra. 
In both these approaches,
the spectrum is modelled as sum of two main components, one arising from
an accretion disk and the other from the boundary layer between the disk and
the neutron star surface. In the first approach (sometimes referred to as the
``Western'' approach), the accretion disk is considered to be a 
standard cold disk
emitting a soft component, while the boundary layer radiates the harder
Comptonized component \citep{Mit84, Mit89, Dis02, Agr03}. In the 
second approach (called the ``Eastern''
approach), the boundary layer emits a cool black body emission while the
harder Comptonized component arises from 
an hot inner disk \citep{Whi86,Dis00,Dis01}. There is also evidence for 
a variable hard power-law component
in the spectra of five Z-sources \citep{Asa94, Dis00, Dis01, Dis02, Dam01}.
The two approaches are difficult to distinguish primarily due to the
absence of good quality low energy ($0.1$-$1.0$ keV) data and the uncertainty
in the absorption column density.

To understand the
phenomenology of these sources, it is important to study their complete
spectral evolution in order to confirm any generic behaviour and to see if
there are sub-classes of sources that have similar spectral evolution.
There is already some evidence that  Z-sources may not be an homogeneous group. 
This was first
indicated by \cite{Kul96,Kul97} who classified them into 'Sco' 
and 'Cyg' like sources.
Detailed study of spectral evolution along the Z-track has been carried out for
GX 349+2 \citep{Agr03}, Cyg X-2 \citep{Dis02}, GX 17+2 \citep{Dis00}
and GX 340+0 \citep{Chu06}. In general, there is considerable variation
of the spectral parameters as a source moves along the Z-track and 
the variations
are not always similar in all sources, indicating that Z sources are not
an heterogeneous group. 
For example, as Cyg X-2 moves from the HB to NB,  the optical depth of
the Comptonized component decreases while the electron temperature increases
dramatically from  $3$ to $8$ keV. A more modest increase in electron 
temperature
from $2.7$ to $3.2$ is seen in GX 349+2. On the other hand, for GX 17+2
the temperature decreases along the same track. Along the flaring branch,
detailed spectral evolution study has only been undertaken for GX 349+2
\citep{Agr03} where the temperature increases along the track. 
While these results indicate that the physical parameters and conditions
for these sources are different, what seems to be generic is that the 
effect of Comptonization decreases
along the normal branch while it increases along the flaring branch. 
This is revealed by
the variation of the Comptonization parameter $y =4(kT_e/m_ec^2) \tau^2$, 
where kT$_e$ is electron temperature, m$_e$ is rest mass of electron, $c$ is speed of light, and $\tau$ is optical depth of corona. The Comptonization parameter $y$  decreases
along the normal branch and increases along the flaring one,  
a behaviour clearly seen in GX 349+2.

\begin{table}
\caption{RXTE observations of LMC X-2}
\begin{tabular}{cllc}
\hline 
\hline 
OBSID             & Start Date   & End Date  & Duration \\
&&& (ksec) \\
\hline
50041-01-01-**    & 2001 Feb 10  & 2001 Feb 14 & 145\\
60027-01-01-** & 2001 Aug 30  &  2001 Sep 3 & 144\\
60027-01-02-** & 2001 Dec  13 & 2001 Dec 16  &122\\
60027-01-03-** & 2002 Feb 01 & 2002 Feb 06 &148\\
70017-01-**-** & 2002 Aug 22 &  2002 Oct 30 &185\\
\hline
\end{tabular}
\label{Obsid}
\end{table}

The temporal properties of Z-sources are also known to correlate
with the position of the source in the Z-track. Three different kinds of
quasi-periodic oscillations (QPOs) have been seen in these sources.
QPO with frequency 15-100 Hz, which appear in the horizontal branch (but sometimes
seen up to the upper part of the normal branch), are called horizontal
branch oscillations (HBO). The frequency of an HBO
increases from top left to bottom
right of the horizontal branch \citep{Hom02, Wij97}.
QPO with frequency 5-15 Hz are observed in the normal and the flaring
branch, and are  called normal-flaring branch
oscillations \citep{Hom02}. These low frequency oscillations have been seen
in all the galactic Z-sources except GX 349+2 where instead 
broad peaked noise is observed \citep{Agr03a, One02}. The third
type of QPO are kHz ones (300-1200 Hz) which are observed in all the branches.
The frequency of kHz QPO increases along the Z-track from HB to FB \citep{Van00}

Early observation of LMC X-2 showed that its luminosity varies from
$0.6$ to $3 \times$ 10$^{38}$ erg s$^{-1}$ \citep{Mar75}. LMC X-2
was identified as a Z-source when it 
traced out a complete Z-track during 2001-2002 RXTE observations
\citep{Sma03}. Its optical counterpart  is a faint blue
star \citep{Pak78} typical of other LMXB. Simultaneous optical and
X-ray observations of this source have revealed that X-ray and optical
emission are correlated with a time delay of $\sim$ 20 s
\citep{Mcg03}. 
A brief ($\sim 10$ ksec) XMM-Newton observation of this source 
can be described by a disk blackbody, with 
inner disk temperature T$_{in}$ $\sim 0.5$ keV and a blackbody component 
with temperature $T_{bb}$ $\sim$ 1.5 keV \citep{Lav08}.  However, the source 
is bright enough to cause pile up in XMM-Newton observations, making such 
spectral analysis difficult. No QPO has been detected in LMC X-2, 
perhaps due to low statistics  data.

\begin{figure*}
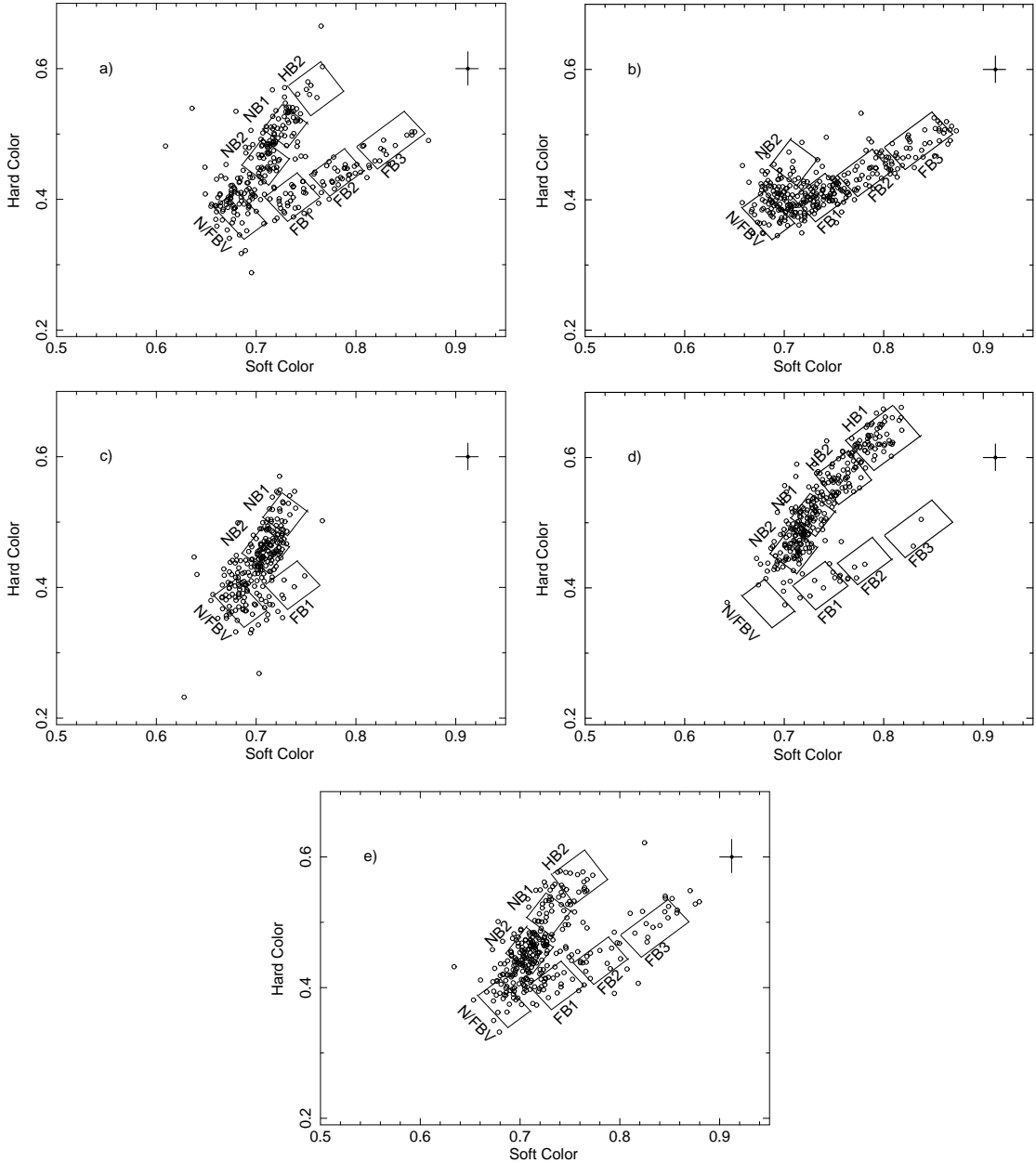

\centering
\begin{tabular}{cc}
\epsfig{file=col_50041.eps,width=0.3\linewidth,angle=-90} &
\epsfig{file=col_60027A.eps,width=0.3\linewidth,angle=-90} \\
\epsfig{file=col_60027B.eps,width=0.3\linewidth,angle=-90} &
\epsfig{file=col_60027C.eps,width=0.3\linewidth,angle=-90} \\
\end{tabular}
\vskip 0.15in
\epsfig{file=col_70017A.eps,width=0.3\linewidth,angle=-90}
\caption{X-ray colour-colour diagram obtained using the RXTE-PCA data
for five observations a) 2001 Feb 10 b) 2001 Aug 30  c) 2001 Dec  13
d)  2002 Feb 01 e) 2002 Aug 22. Soft colour is ratio of count rates in
the 4.5-6.5 and 2.5-4.5 keV energy bands and hard colour is
ratio of count rates in the 9.8-18.5 and 6.5-9.8
keV bands. The colour
values have been calculated using 512 seconds averages. The boxes mark
the regions for which spectral analysis was undertaken. The
representative error-bar for colour values is given in upper right
corner of  each figures}
\label{eachCD}
\end{figure*}

\begin{figure*}
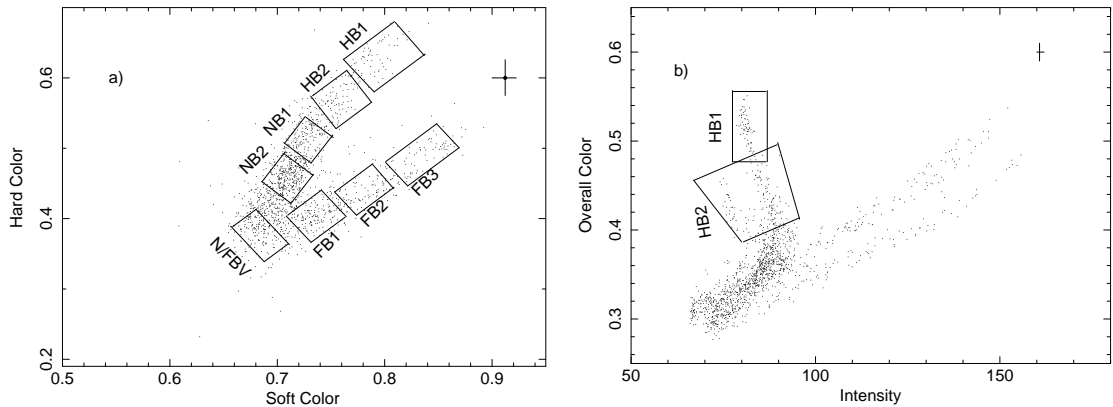

\centering
\begin{tabular}{cc}
\epsfig{file=col_total_new.eps,width=0.3\linewidth,angle=-90} &
\epsfig{file=hid_total_new.eps,width=0.3\linewidth,angle=-90} \\
\end{tabular}
\caption{a) A single, combined colour-colour diagram for the  five different
observations (see Table \ref{Obsid}). The energy bands and time bin size used to
construct the combined colour-colour diagram is same as that used in Fig. \ref{eachCD}.
b) Combined hardness-intensity diagram for five different
observations.  The overall colour is the ratio of count rates in the
6.5-18.5 and 2.5-6.5 keV energy ranges. The intensity is for 2.5-18.5
keV and is in units of counts/sec/2PCUs. 
Each points corresponds to 512 sec averages. The
representative error bar for each data points is given in the upper
right corner of each figure.}
\label{CDtot}
\end{figure*}

LMC X-2 provides an unique opportunity to understand the spectral evolution
of Z sources and the radiative processes that occur in such sources.
The distance ($\sim 50\pm 2$ kpc) to the source  \citep{freed01,mc03} is much less uncertain 
than those of galactic sources enabling a more accurate estimation of
its intrinsic luminosity. The Galactic column density along the 
line of the source is $6.3 \times 10^{20}$ cm$^{-2}$. This provides a
firm lower limit to the absorption  which 
can provide constrains on spectral models. The source has been extensively
observed by the {\it Rossi X-ray Timing Experiment} (RXTE) for a good time
duration of $\sim 750$ ksec. This allows
the construction of elaborate CCD and HID diagrams and a detailed study of
its spectral evolution. In addition, a $\sim 50$ ksec 
observation by {\it Suzaku} reveals the broad band spectrum ($0.3$-$30$ keV) 
of this source  which allows models to be differentiated and the 
bolometric luminosity to be estimated.

In this paper, we report the results of spectral analysis of RXTE in \S 2 and
{\it Suzaku} data in \S 3. In \S 4, we summarise the main results and discuss
the implications.

\section{RXTE OBSERVATIONS}
\subsection{Colour-Colour and Hardness Intensity Diagrams}
LMC X-2 was observed by RXTE during 2001-2002 for a total good time of
750 ksec. The Observation log is shown in Table \ref{Obsid}. We have analysed
the data collected by Proportional-Counter-Array (PCA) which operates
in 2-60 keV energy band and consists of five proportional counter
units with a total effective area of $\sim$ 6500 cm$^2$
\citep{Jah96}. The analysis was done using PCA
standard-2 mode data. Since only two of the five proportional counter
units (PCU0 and PCU2) were reliably on all through the observation,
only data from these PCUs were selected for analysis.  We define soft
colour as ratio of count rates in the 4.5-6.5 keV and 2.5-4.5 keV
energy bands and hard colour as that in the 9.8-18.5 keV and 6.5-9.8
keV bands. The CD, generated using 512 sec average counts, for different
observations are plotted in Figure \ref{eachCD}. In Figure \ref{CDtot}(a), the
combined CD for all the observations is plotted. The 
result is similar to that obtained by \cite{Sma03}, except that the
present analysis is for a larger data set. We define overall
colour as ratio of count rates in the 6.5-18.5 keV and 2.5-6.5 keV
bands. The Hardness Intensity Diagram (HID) was constructed by 
plotting the overall colour against the 2.5-18.5 keV counts/sec/2PCUs and
is plotted in Figure \ref{CDtot}(b). 
 Since there was no gain
changes after 2000 there was no need to normalise the intensities and
colours.

As seen in Fig \ref{CDtot}(a), the horizontal branch (HB) for this
source is neither horizontal nor distinct. The classification requires
study of other spectral relationships like HID which is shown in Fig \ref{CDtot}(b). 
 The HID of LMC X-2 is quite similar to that
seen for another well studied Z-source GX 17+2 (see Fig.1 of \citealt{Hom02}). 
In the HID of GX 17+2, the slanted branch (almost vertical)
at the top left is called the horizontal branch and 
we do a similar identification
for LMC X-2. We divide the HB in the HID,  into two region HB1 and HB2 and 
mark the corresponding location in the CD. The normal and flaring branch are
distinct in the CD. Thus, we are able to identify all three branches for
LMC X-2. For spectral studies, the Z-curve has been divided into eight regions.
Two for the horizontal branch (HB1 and HB2), two for the normal branch
(NB1 and NB2), a normal/flaring branch vortex point (N/FBV) and three
regions for the flaring branch (FB1, FB2 and FB3). All eight regions
have been marked out in Fig \ref{CDtot}(a).

\subsection{Spectral Evolution}

We extracted spectra corresponding for the eight regions in the CD to
carry out a a detailed spectral evolution study.  We added 1 \%
systematic error to all the spectra to take in to account the uncertainty
in the PCA response matrix.

\begin{figure}
\begin{center}
\includegraphics[height=1.0\linewidth, angle=-90]{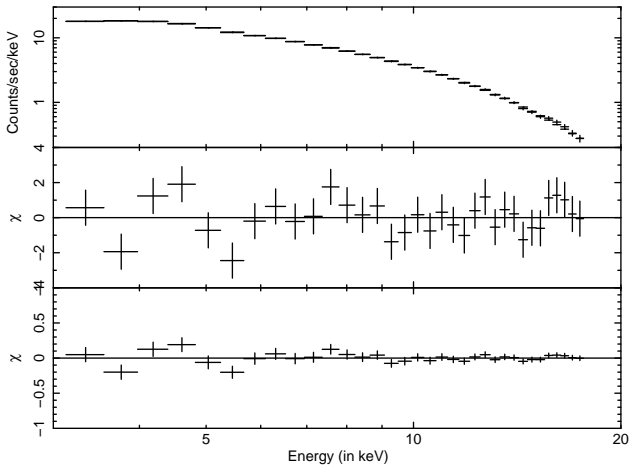}
\end{center}
\caption{The observed PCA spectrum of LMC X-2 in the HB together with
the best fit diskbb+compTT model are shown in panel 1. Residuals in
unit of sigma  without and with inclusion of systematic errors are
shown in panel 2  and panel 3 respectively.}.
\label{PCAspec}
\end{figure}

 A single component Comptonization model (compTT in XSPEC; see
\citealt{Tit94}) with absorption provided adequate fit to all the PCA
spectra in the energy range $3$-$20$ keV. Since the data is for
energies $> 3 $ keV,  the absorbing column density is not well
constrained. Hence we adopt a constant value of N$_H \sim 9 \times 10^{20}$ 
based on the {\it Suzaku} analysis reported in section 3.  An
iron emission line centred at $\sim$ 6.4 is also required to improve
the fit.  In the top panel of Fig.\ref{PCAspec} we show the observed count rate
spectrum for HB1, along with the best fit model.  The residuals (in
units of $\sigma$), with and without inclusion of systematic error are
shown in the middle and bottom panels respectively. The PCA data
suggests that the soft component, originating either from the disk
or from the boundary layer, is absent above 3 keV.

The spectral parameters of compTT component evolve significantly as the
 source moves along the Z-track. The best fit parameters at different
 parts of Z-track have been listed in Table \ref{specpar} and evolution
 along the track has been plotted in Fig. \ref{spec_var}. The electron temperature
 decreases as the source evolves from HB to lower part of FB. This
 variation is different from that seen in GX 349+2 \citep{Agr03} and Cyg
 X-2 \citep{Dis02} where the opposite behaviour was found. On the other
 hand it is similar to that observed in GX 17+2 \citep{Dis00}. As LMC
 X-2 moves along the flaring branch the temperature increases which
 again is in contrast to the behaviour observed for GX 349+2 where the
 temperature decreases. However what seems to generic is that effect
 of Comptonization decreases along the normal branch while it increases
 along the flaring branch. This is revealed by variation in Comptonization
 parameter y=4(kT$_e$/m$_e$c$^2$)$\tau^2$ which decreases along the normal branch
 and increases along the flaring one (Fig. \ref{spec_var}) a behaviour also seen in
 GX 349+2. In the flaring branch as $y$ increases the spectrum is that
 of saturated Comptonization with peak $\sim$ $3kT_e$ $\sim$ 6 keV.

\begin{table*}
\caption{The best fit parameters obtained by fitting the RXTE-PCA data
with XSPEC model compTT. The electron temperature $kT_e$ and seed
photon temperature $kT_W$ is measured in keV. $\tau$ is optical depth
of central corona and $y$ is Compton y-parameter. The division N/FBV
means normal-flaring branch vertex}
\begin{tabular}{ccccccccc}
\hline
\hline
   & HB1  & HB2  & NB1  & NB2  &N/FBV  & FB1  & FB2 & FB3\\
\hline
$kT_W$ & 0.47$^{+0.03}_{-0.06}$ & 0.55$^{+0.04}_{-0.06}$ & 0.61$^{+0.03}_{-0.04}$ & 0.62$^{+0.03}_{-0.03}$ & 0.59$^{+0.03}_{-0.04}$ & 0.61$^{+0.03}_{-0.03}$ & 0.62$^{+0.03}_{-0.04}$ & 0.58$^{+0.04}_{-0.05}$ \\ 
& & & & & & & \\
$KT_e$ & 2.74$^{+0.04}_{-0.04}$ & 2.60$^{+0.04}_{-0.04}$ & 2.45$^{+0.03}_{-0.03}$ & 2.24$^{+0.02}_{-0.02}$ & 1.95$^{+0.03}_{-0.02}$ & 1.94$^{+0.02}_{-0.02}$ & 2.00$^{+0.02}_{-0.02}$ & 2.07$^{+0.02}_{-0.02}$ \\
& & & & & & & \\
$\tau$ & 13.37$^{+0.22}_{-0.24}$ & 12.90$^{+0.25}_{-0.26}$ & 12.59$^{+0.25}_{-0.25}$ & 12.99$^{+0.25}_{-0.25}$ & 14.47$^{+0.33}_{-0.34}$ & 15.71$^{+0.40}_{-0.40}$ & 16.51$^{+0.45}_{-0.40}$ & 18.13$^{+0.41}_{-0.31}$ \\
& & & & & & & \\
$y$ & 3.83$^{+0.14}_{-0.15}$ & 3.38$^{+0.14}_{-0.15}$ & 3.04$^{+0.13}_{-0.13}$ & 2.96$^{+0.12}_{-0.12}$ & 3.19$^{+0.15}_{-0.15}$ & 3.74$^{+0.19}_{-0.19}$ & 4.26$^{+0.24}_{-0.21}$ & 5.32$^{+0.25}_{-0.19}$ \\
& & & & & & & \\
$\chi^2$/dof & 17.7/29 & 17.2/29 & 15.5/29 & 12.7/29 & 30.1/29 & 15.6/29 & 16.2/29 & 19/29 \\
\hline
\end{tabular}
\label{specpar}
\end{table*}

\section{SUZAKU OBSERVATION}
\subsection{Colour-Colour Diagram and Light curves}
{\it Suzaku} observed LMC X-2 for a good time duration of 119 ksec
on 2006 April 24 and 25. 
The observatory \citep{Mit07} consists
of two different types of instrument, the X-ray Imaging Spectrometer
(XIS; \citealt{Koy07}) and Hard X-ray Detector (HXD; \citealt{Tak07} ).

 There are four XIS units (named XIS0, XIS1, XIS2, XIS3), each
of them is a 1024 $\times$ 1024 pixel CCD and covers an energy range of
0.3-12 keV. HXD consists of two types of
detectors, the Si diode detector (PIN) and the Gadolinium silicate crystal (GSO). 
 The PIN diode covers an energy range of 10-70 keV while the GSO scintillators are
sensitive in 40-600 keV band.  The XIS data were collected in normal mode
with 1/4 window option. The minimum bin time available in this mode is 2
seconds. The cleaned XIS event files were used to obtain the light-curves
and spectra. The XIS image is shown in Fig. \ref{Suz_img}. 
The source light-curves were extracted using 4$\arcmin$ radius centred at source position.

\begin{figure}
\begin{center}
\includegraphics[width=1.0\linewidth,angle=0]{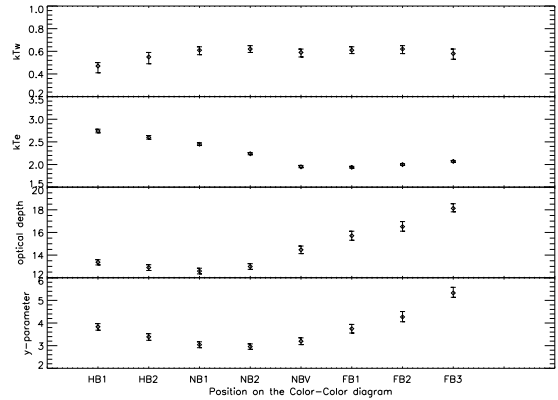}
\end{center}
\caption{The variation in spectral parameters obtained by fitting the
PCA data with Comptonization model. The main variation is that of the
temperature of Comptonization cloud which decreases from the upper
part of the HB to the lower part of the NB. The Compton y-parameter
also show a similar variation}
\label{spec_var}
\end{figure}

\begin{figure}
\begin{center}
\includegraphics[height=0.8\linewidth]{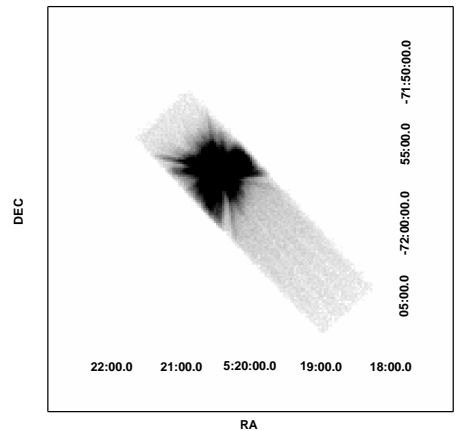}
\end{center}
\caption{The XIS image of LMC X-2 }.
\label{Suz_img}
\end{figure}

\begin{figure}
\begin{center}
\includegraphics[height=1.0\linewidth, angle=-90]{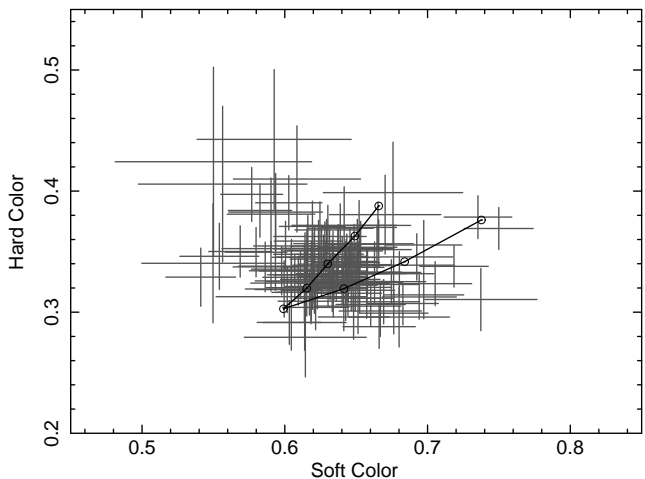}
\end{center}
\caption{The colour-colour diagram constructed using XIS0 data. The hard
colour is ratio of count rates in 6.5-10 keV and 4.5-6.5 keV energy
bands, while the soft colour is ratio of count rates in the energy
bands 4.5-6.5 keV and 3.0-4.5 keV. The solid line connects the eight
simulated colours on the parameters corresponding to RXTE observations
(Table \ref{specpar}).}
\label{suz_col}
\end{figure}

To compare with RXTE observations, we study the colour and intensity
properties of the counts at energies $>$ 3 keV. Fig. \ref{suz_col} shows the
colour-colour diagram for the {\it Suzaku} data using three energy
bands defined as 3-4.5 keV, 4.5-6.5 keV and 6.5-10 keV. The data
points corresponds to 512 sec averages. Superimposed on the figure
is the simulated colour-colour diagram, obtained using the best fit 
PCA spectral parameters and XIS0 response matrix. 
In particular, eight  fake X-ray spectra were created using the best fit
RXTE model parameters (Table \ref{specpar}) using the Suzaku 
XIS response matrix. The count rates in
 three energy bands (3-4.5, 4.5-6.5, 6.5-10 keV) were calculated  for these
eight simulated spectra and the simulated colour-colour diagram was computed. 
The simulated CD is represented by a solid line connecting the eight 
points.  While it is difficult to differentiate the different branches,
the {\it Suzaku} data points are consistent with simulated ones. Note that
statistical error for these data points are significantly larger than for
RXTE data because the 3-10 keV count rates for Suzaku ($\sim$ 6 counts/sec) is
smaller than that for RXTE, $\sim$ 60-150 counts/sec/2PCUs. Comparison with
hardness intensity plot generated using RXTE data is more difficult
because apart from the statistical uncertainty of {\it Suzaku} points,
the absolute calibration between the PCA and {\it Suzaku} may differ.

\begin{figure}
\begin{center}
\includegraphics[height=1.0\linewidth, angle=-90]{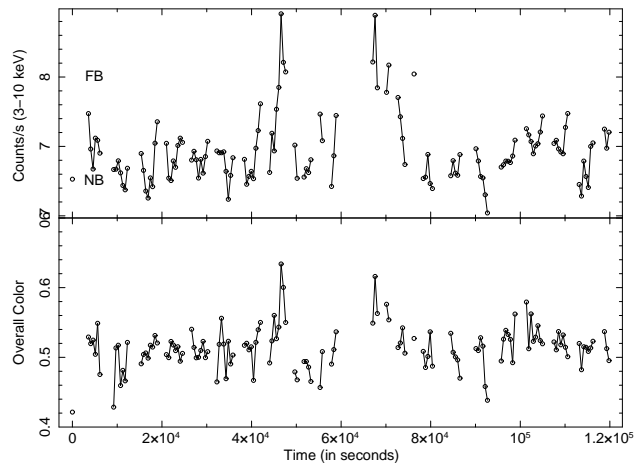}
\end{center}
\caption{Top panel: The XIS0 light curve where the intensity is the
count rate in 3-10 keV energy band. Bottom panel: overall all colour
defined as ratio of the count rates in the 5-10 keV and 3-5 keV energy
bands as function of time. The source hardens during the flares.}
\label{suz_int}
\end{figure}

The XIS0 light-curve created using 3-10 keV energy band is shown in the
top panel of Fig. \ref{suz_int}. There are clear flaring like activity where the rate
increases by $\sim$ 30 \%. Moreover the spectra hardens during the flares as
revealed in the bottom panel of Fig. \ref{suz_int} where overall colour is plotted
as function of time. The overall colour is ratio of count rates in
5-10 and 3-5 keV energy bands. Thus, we divided the data in two states,
a flaring state where the count rate is $>$ 7 counts/sec and a non flaring
one corresponding to rates $<$ 7 counts/s. The low statistics of the data does not allow
for more reliable divisions. We identify the flaring state with the flaring
branch and the low count rate state as the normal branch.

\subsection{Spectral Evolution}
We extracted source and background spectra for the flaring and non flaring
parts  for  each of the four XIS units. The latest calibration database was used to create the
XIS response. The response matrix files were generated using FTOOLS task
{\it xisrmfgen} and ancillary response files were generated using the
FTOOLS task {\it xissimarfgen}.

To analyse the HXD data first we checked for the process
version. Since the HXD process version was 2.0.6.13, we reprocessed
the data as recommended by the {\it Suzaku} team. As a first step to
reduction, we ran the command {\it hxdtime} to calculate HXD event
arrival time correction. Then invariant pulse-heights (PI) were
determined using {\it hxdpi}. The HXD event grades were calculated
using the task {\it hxdgrade}. We used standard screening criteria
(given in {\it Suzaku} ABC guide) to obtain the filtered PIN event
files. We also excluded the telemetry saturation period while creating
filtered event files. The filtered PIN files were used to create 
the source spectra for the different flux level and dead time
correction was performed to the PIN spectra. As suggested by the HXD
team, we used version 1.2 (Method = LCFIT(bgd\_d)) background file to
create PIN background spectra. Since the PIN background was simulated
with a ten times scaled level to decrease the Poisson noise, we have
modified exposure keyword by entering a new exposure ten times the
original. Since we do not have enough counts above 30 keV, we
restricted our analysis to $<$ 30 keV and did not use GSO data. We
added the data, background and response matrices from the FI (front
illuminated) XIS CCDs. We used 0.5-10 keV  FI CCDs data, 0.3-10 keV
BI (back illuminated) CCD data and 10-30 keV for PIN HXD for combined
spectral fitting. We grouped the XIS data to get 3 channel per
resolution and added 2 \% systmatic error to take into account the
uncertainty in the XIS response matrix. While spectral fitting, we allow for
calibration uncertainties between FI CCDs, BI CCDs and PIN HXD.

\begin{table}
\caption{The best fit parameter obtained by fitting {\it Suzaku} data with black body (BBODY), Comptonization (COMPTT) and 
broad Gaussian Iron line models. $kT_{BB}$ is the blackbody temperature. $kT_W$, $kT_e$ and $\tau$ are the seed photon
temperature, the electron temperature and the optical depth of the Comptonization model. $E_{Fe}$ and $\sigma_{Fe}$ are the
centroid energy and the width of the Iron line Gaussian.}
\begin{tabular}{lll}
\hline
Para.   &   NB  & FB \\
\hline
\hline
$N_H^a$($ 10^{20}$ cm$^{-2}$) & $<6.3$ & $<6.3$ \\
&           &  \\
$kT_{BB}$ (eV) & 213$^{+2}_{-3}$ & 215$^{+2}_{-1}$ \\
&           &  \\
$kT_W$ (eV) & 528$^{+3}_{-3}$ & 530$^{+4}_{-4}$ \\
&           &  \\
$kT_e $ (eV) & 2180$^{+20}_{-6}$ & 2190$^{+20}_{-8}$ \\
&           &  \\
$\tau$ & 14.05$^{+0.04}_{-0.04}$ & 14.75$^{+0.04}_{-0.20}$ \\
 &   & \\
 $E_{Fe} $ (keV) & 6.4 (fixed)    &     6.4 (fixed)  \\
 & & \\
$\sigma_{Fe}$ (keV) & 0.27$^{+0.19}_{-0.14}$ & 0.27 (fixed)  \\
& &  \\
EqWidth$^b$  (eV) & 20  & 9  \\
& &  \\
Ftest Prob$^c$ & 7 $\times$ 10$^{-3}$   & 0.11 \\
& &  \\
$F_{line}$$^d$ & 0.74$\pm$0.35 & $<0.9$$^e$ \\

$\chi^2$/dof & 463/375 & 442/374 \\
\hline
\end{tabular}

\label{suz_west_par}
$^a$The column density $N_H$ was restricted to be larger than the Galactic value of 6.3 $\times$ 10$^{20}$ cm$^{-2}$. 
$^b$The Equivalent Width of the Iron line. $^c$The F-test probability for the presence of the Gaussian Iron line. $^d$The flux of iron line in unit of 10$^{-12}$ erg/s/cm$^2$.
$^e$ 90\% confidence upper limit on the iron line flux
\end{table}

\begin{table}
\caption{The best fit parameter obtained by fitting {\it Suzaku} data with disk black body (DISKBB), Comptonization (COMPTT) and 
broad Gaussian Iron line models.  $kT_{in}$ and $R_{dbb}$ are the inner disk temperature and radius. 
$kT_W$, $kT_e$ and $\tau$ are the seed photon
temperature, the electron temperature and the optical depth of the Comptonization model. $E_{Fe}$ and $\sigma_{Fe}$ are the
centroid energy and the width of the Iron line Gaussian. The luminosities are computed using the best fit model in the 
energy range 0.01-50 keV and are in units of $10^{38}$ ergs/sec.}
\begin{tabular}{lll}
\hline
Para.   &   NB  & FB \\
\hline
\hline
$N_H$ ($\times$ 10$^{20}$ cm$^{-2}$) & 8.6$^{+0.2}_{-1.0}$ & 8.35$^{+0.42}_{-1.1}$\\
&           &  \\
$kT_{in}$ (eV) & 394$^{+2}_{-1}$ & 397$^{+13}_{-14}$ \\
&  & \\
$R_{dbb}$ (km) & 99$\pm$3 & 95$\pm$5 \\
&           &  \\
$kT_W$ (eV)  & 660 $\pm$20 & 670$\pm$20 \\
&           &  \\
$kT_e$ (eV) & 2600$^{+10}_{-40}$ & 2490$^{+10}_{-50}$ \\
&           &  \\
$\tau$   & 11.30$^{+0.20}_{-0.03}$ & 12.40$^{+0.15}_{-0.15}$ \\
&           &  \\
$E_{Fe} $  (keV) & 6.4 (fixed)  & 6.4 (fixed) \\
   & & \\ 
$\sigma_{Fe}$ (keV) & 0.56$^{+0.15}_{-0.12}$ & 0.44$^{+0.11}_{-0.10}$  \\
&  &  \\
EqWidth$^a$ (eV) & 72  & 36                    \\
&  & \\
Ftest Prob$^b$ & 7.1 $\times$ 10$^{-9}$ &  2.5 $\times$ 10$^{-4}$\\
&           &  \\
$L_{abs}^c$ & 1.83$\pm$0.11 & 1.95$\pm$0.12 \\
&           &  \\
$L_T^d$ & 2.10$\pm$0.13 & 2.23$\pm$0.12 \\
&           &  \\
$L^e_{dbb}$ & 0.59$\pm$0.04 & 0.60$\pm$0.04 \\
&           &  \\
$L_{compTT}$ & 1.51$\pm$0.03 & 1.63$\pm$0.03 \\
&           &  \\
$y^g$ & 2.60$^{+0.09}_{-0.04}$ & 2.98$^{+0.07}_{-0.09}$ \\
&           &  \\
$F_{line}^h$ &$2.68\pm$0.54 & 1.73$^{+1.42}_{-1.04}$\\
$\chi^2$/dof & 380/375 & 356/373 \\
\hline
\end{tabular}

$^a$The Equivalent Width of the Iron line. $^b$The F-test probability for the presence of the Gaussian Iron line.
$^c$Total absorbed luminosity. $^d$Total unabsorbed luminosity. $^e$Unabsorbed luminosity of disk black body component.
$^f$Unabsorbed luminosity of Comptonized component. $^g$The Comptonization parameter. $^h$The flux of iron line in unit of 10$^{-12}$ ergs/s/cm$^2$
\label{suz_east_par}
\end{table}

The broad band spectra of {\it Suzaku} allows for a two component fit
to the data. We use a blackbody and Comptonization model (XSPEC model
COMPTT) to represent the ``Western'' model and fit both the flaring
and normal branch data sets. A Gaussian representing iron line emission is
statistically significant for the normal branch. Since the Galactic neutral hydrogen
column density in the direction of LMC X-2 is 6.3 $\times$ 10$^{20}$
cm$^{-2}$, we impose this as a lower limit to $N_H$. The results of
spectral fits are shown in Table \ref{suz_west_par}. The best fit $N_H$ gets pegged to
the lower limit for both the data sets. The spectral variation is
entirely due changes to Comptonization.

We repeat the spectral fitting using disk black body (XSPEC model DISKBB)
and the same Comptonization model to represent ``Eastern'' model. The
best fit spectral parameters are shown in Table \ref{suz_east_par}. Here, the column
density turns out to be $\sim$ 9 $\times$ 10$^{20}$ cm$^{-2}$ above the
galactic lower limit. For the normal branch $\chi^2$/d.o.f = 380/375
which is significantly better than the one obtained for the ``Western''
model $\chi^2/d.o.f$ = 463/375. A similar trend is seen for the flaring
branch, where the ``Eastern'' model gives better $\chi^2$/d.o.f =
356/373 as compared to $\chi^2$/d.o.f = 442/374. Thus spectral fitting
of {\it Suzaku} data strongly prefers the interpretation where the
soft component is disk emission rather than simple black body. Note
that this distinction is possible because of the imposed lower limit on
the absorption column density. If the column density is allowed to have any
value then a reasonable fit using the black body component is possible. For the model
with black body emission,  the best fit column density turns out to be $N_H = $ 0.013 $times$ 10$^{22}$~cm$^{-2}$ with a  $\chi^2$/dof = 367/375.  For
Galactic sources, the column density is more uncertain and hence such
distinction would not have been possible even with broad band data.

\begin{figure}
\begin{center}
\includegraphics[height=1.0\linewidth, angle=-90]{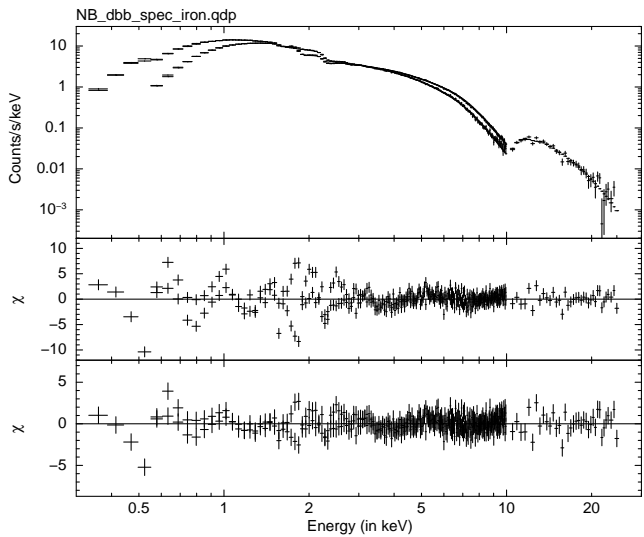}
\end{center}
\caption{The top panel shows the observed {\it Suzaku} spectrum of LMC
X-2 for the normal branch along with best fit disk blackbody and
Comptonization model. The lower panels show the residuals with (bottom
panel) and without (middle panel) systematic error addition of 2\%.}
\label{suz_spec}
\end{figure}

The spectral fit to the non-flaring  state is shown in Fig. \ref{suz_spec}. Also plotted
are the residuals with and without the addition of a systematic error
of 2\%. The column density is well constrained to 9 $\times$ 10$^{20}$
cm$^{-2}$ and it is this value that has been used for fitting RXTE
data (Table \ref{specpar}). The inner disk temperature $kT_{in}$, and the inner
disk radius $R_{in}$ as well as the soft photon input temperature
$kT_W$ remain invariant between the flaring and normal branches. A broad Iron line
is significantly detected for both branches. The
electron temperature decreases during the flare, which is consistent
with behaviour seen by RXTE, although {\it Suzaku} data requires an
higher temperature. More significantly it is the Compton y-parameter
which is seen to increase during the flare, which is again consistent
with the RXTE observations. Using a distance of 50 kpc, the unabsorbed
luminosities of flaring and normal branch turn out to be 2.1 $\times$
10$^{38}$ ergs/s and 2.2 $\times$ 10$^{38}$ ergs/s, which for a 
$1.4 M_\odot$ neutron star correspond to 1.1 and 1.15 of the 
Eddington value.

\section{DISCUSSION}

A detailed spectral analysis of the extragalactic source LMC X-2, has
been undertaken using broad band data from {\it Suzaku} and from a
large ($\sim$ 750 ksec) data set obtained using RXTE-PCA.  

From the
PCA data the complete Z-track evolution was obtained and spectra at
eight representatives position were analysed. In the energy band of the
PCA (3-20 keV), a single component Comptonization fit adequately
represents the spectra from all the track branches. This is unlike
other Z-sources (GX 349+2: \citealt{Agr03}; Cyg X-2: \citealt{Dis02}),
where an additional soft
component is required in the RXTE  energy band. The temperature of
Comptonizing cloud, $kT_e$ is found to decrease from the horizontal to
the normal branch (see Fig. \ref{spec_var}) which is similar to what is observed
for GX 17+2 \citep{Dis00}, but is in contrast to results from 
Cyg X-2 \citep{Dis02} and GX 349+2 \citep{Agr03} where the opposite
behaviour was found. Along the Flaring branch, the temperature
increases for LMC X-2, which is again different from GX 349+2 where
the temperature decreases. On the other hand, for LMC X-2 the Compton
y-parameter decreases along the horizontal and normal branch and
increases along the flaring branch. This is consistent with what is
observed for GX 349+2 and it seems to be a generic feature of Z-sources. 

These results confirm that while details of the spectral evolution are
different for each source, there are generic behaviour like the
evolution of Compton y-parameter which seems to be similar in all
Z-sources. The hard component evolution similarity between LMC X-2 and
GX 17+2 suggests that there could be sub groups within Z sources where
similar physical conditions are met. The observed similarity in the
spectral evolution seen in these two sources is also suggested in the
near identical colour and hardness intensity diagrams of these
sources. Similar diagram are also observed in Sco X-1.  Hence, our
results are consistent with the idea that Z sources can be further
divided into two subclasses, 'Sco'  and 'Cyg' like sources \citep{Kul96,
Kul97} with LMC X-2 being a member of the 'Sco' like class.   
While it was  suggested earlier that the difference between 
'Sco' and 'Cyg' like sources may be due to inclination angle \citep{Kul96}, 
the  spectral differences  between LMC X-2 and other sources, suggest instead
a more intrinsic difference between the two classes. Analysis of Z source XTE J1701-462 reveales that the source evolves from 'Cyg' like to 'Sco' like Z-source, suggesting that diffence between these two classes is not caused by difference in the inclination angle \citep{lin09, Hom07} .

The Suzaku data provides broad band spectral and luminosity evolution
of the source from the NB  to the FB. As expected the data requires an
additional soft component. This additional component was represented
as a black body (emitted from boundary layer) and as emission from
multicolour disk. The multi-colour disk approach (sometimes called the
``Eastern'' approach) provides significantly better fit to the
data. This spectral distinction is possible because the Galactic
neutral hydrogen column density along the direction of the source is
$N_H$ = 6.3 $\times$ 10$^{20}$ cm$^{-2}$ and hence the fitted
absorption column density was restricted to be larger than this
value. Such a constrain is usually not available for Galactic sources.

The distance to LMC X-2 (50$\pm$2 kpc) is better constrained than those of
Galactic sources and hence broad band {\it Suzaku} data provides a
better estimate of the bolometric luminosity. The unabsorbed
luminosity of the source for the normal branch is 2.1 $\times$
10$^{38}$ erg/s and for the flaring branch is $\sim$ 2.23$\times$
10$^{38}$ erg/s. Taking into account the uncertainty in the spectral
fitting ($\sim$ 5\%), the distance($\sim$ 5 \%) and calibration of
different instrument ($<$ 10 \%), one can put a highly conservative 
error estimate of $ < 15$\% on these luminosities. If the source reaches its Eddington limit on
the normal branch, this would imply that the mass of neutron star is
1.50$\pm$0.1 M$_\odot$. If instead the source is Eddington limited in
the flaring branch one obtains a slightly higher mass of 1.61$\pm$0.09
M$_\odot$.

The phenomenological spectral models used in this analysis, a single temperature
Comptonized component and a multi-colour disk emission, is adequate considering
the statistics of the available data and the uncertainties regarding theoretical models.
However, considering that the system is a near Eddington neutron star source, it is
expected that the actual radiative processes would be considerably more complicated
with possibly quantitative and qualitative differences as compared to the simple model
adopted here. For example, detailed structural analysis of the boundary layer
reveals a temperature and density stratified extended region \citep{Pop01}.

A fairly detailed model for Z sources is where
soft photons produced  by electron cyclotron emission in the neutron star magnetosphere are
Comptonized primarily in a hot central corona \citep{Psa95,Psa97}. The outer accretion disk
gets converted into a radial flow which also  Comptonises the outgoing photons. The phenomenological
description in this analysis is broadly consistent with this picture with the temperature
and optical depth of the Comptonizing component identified as the average temperature
and optical depth of the hot central corona. The source of seed photon for Comptonization
which is assumed to be a black body with temperature, $kT_W$ in the XSPEC model COMPTT,
will then be the electron cyclotron emission. Finally the multi-colour disk component is the
outer disk. Note that from Table \ref{suz_east_par}, the seed photon temperature, $kT_W \sim 0.65$ keV is different
from the inner disk temperature, $kT_{in} \sim 0.4$ keV and hence in this interpretation the
seed photons are not produced by the outer disk. This is consistent with the above model,
where the seed photons are produced in the magnetosphere.

A straightforward interpretation of the analysis is that
the inner disk radius, $R_{dbb}$ does not vary significantly when the source is in a flaring
state and at $\sim 100$ km is about ten times the neutron star radius. The radius may be
larger if the colour factor is significantly greater than unity as is assumed for the radius
estimation. The luminosity of the disk emission is nearly 25\% of the total
luminosity and hence gravitational energy dissipation is not sufficient to power the disk.
At ten times the star radius, the gravitational energy dissipated should be only 10\% or less
of the total dissipation. Instead the disk is probably heated by the inner corona and hence
radiates such high luminosities. This is expected if the disk is truncated due to the
radiation pressure of a near Eddington inner corona. However, contrary to what is seen, in such a scenario,
one would expect the inner disk radius and its luminosity should vary as the source goes into
a flaring mode. The large inferred inner disk radius argues against the possibility that the
broad Iron line detected arises from the disk. Instead, it may be speculated that
it arises perhaps from the neutron star surface. Unfortunately, the line is weak (with
equivalent width of $\sim 50$ eV) and hence the statistics is not sufficient to test
these ideas by modelling the line profile.

It may also be that the phenomenological models used in this analysis do not sufficiently
approximate the complex radiative process of the system. In particular, the soft seed photon
input is assumed here to be a black body and this may not be the case. While the
high energy spectrum is largely independent of the shape of the seed photon spectrum,
the low energy spectrum will depend on its shape. In the framework of a model, the
magnetosphere emission could be more complex than a black body. Moreover there
could be additional source of seed photons from the outer disk. Such complexities
may vary the disk emission parameters or even eliminate the need for the component. 
However, an arbitrary seed photon spectral shape will not be falsifiable by spectral
analysis, since some shape will always be a good representation of the data. A theoretically
motivated spectral shape for the seed photon is required. Despite these caveats, the
spectral analysis shown in this work does reveal the overall behaviour of the source.
Model independently one can still conclude that the lower energy part of the
spectrum (which in this case is the seed photon source and the disk component)
do not vary as the source flares and the entire variation can be ascribed to the
Comptonizing cloud.

The low statistics of the {\it Suzaku} data did not allow for a detailed
broad band spectral study of the source at it evolves along the Z track.
Instead the data was split into two broad parts representing the flaring
and non flaring periods.  This is unfortunate, because such a broad band evolution
study would have been able to test whether the bolometric luminosity increases
monotonically along the track as expected by theoretical models and inferred
by UV observations.  The two main flares observed by {\it Suzaku} seem to be
of similar strength (Fig \ref{suz_int}), however, clearly this needs to be confirmed
by longer duration observations. This will have important implications on 
whether the source is Eddington limited during the flares or not.
These deficiencies may be alleviated by a longer duration {\it Suzaku} observations undertaken
simultaneously with RXTE. The PCA data would distinguish the different track
positions, while {\it Suzaku} would provide broad band spectra for each
position. Alternatively, such results can be obtained by a long duration
observation of the source by the forthcoming multi-wavelength satellite,
ASTROSAT. The proportional counter array on board (LAXPC) would determine
the Z-track, while the UV and other X-ray instruments would give simultaneous
broad band coverage.


\begin{thebibliography}{99}
\bibitem[\protect\citeauthoryear{Agrawal and Sreekumar}{2003}]{Agr03}
Agrawal V.K., Sreekumar P., 2003, MNRAS, 346, 933
\bibitem[\protect\citeauthoryear{Agrawal and Bhattacharyya}{2003}]{Agr03a} 
Agrawal V.K., Bhattacharyya S., 2003, A\&A, 398, 223
\bibitem[\protect\citeauthoryear{Asai et al.}{1994}]{Asa94} 
Asai K. et al., 1994, PASJ, 46, 479
\bibitem[\protect\citeauthoryear{Barnard et al.}{2003}]{Bar03}
 Barnard R., Kolb U., Osborne J.P., 2003, A\&A, 411, 553
\bibitem[\protect\citeauthoryear{Church, Halai \& Blaucinska-Church}{2006}]{Chu06} 
Church M.J., Halai G.S., Blaucinska-Church M.,  2006, A\&A, 460, 233
 \bibitem[\protect\citeauthoryear{D'Amico et al. }{2001}]{Dam01}
 D'Amico F., Heindl W.A., Rothschild R.E., Gruber D.E., 2001, ApJ,  547, L147
 \bibitem[\protect\citeauthoryear{Di Salvo et al. }{2000}]{Dis00} 
Di Salvo T. et al., 2000, ApJ, 544, L119
 \bibitem[\protect\citeauthoryear{Di Salvo et al. }{2001}]{Dis01}
 Di Salvo T. et al., 2001, ApJ, 554, 49
\bibitem[\protect\citeauthoryear{Di Salvo et al. }{2002}]{Dis02} 
Di Salvo T. et al., 2002, A\&A, 386, 535
 
\bibitem[\protect\citeauthoryear{Freedman et al. }{2001}]{freed01} 
Freedman W.L., Madore B.F., Gibson B.K. et al., 2001, ApJ, 553, 47

\bibitem[\protect\citeauthoryear{Hasinger and van der Klis}{1989}]{Has89} 
Hasinger G., van der klis M., 1989, A\&A, 225, 79
\bibitem[\protect\citeauthoryear{Hasinger et al.}{1990}]{Has90}
Hasinger G., van der klis M., Ebisawa K., Dotani T., Mitsuda K., 1990, A\&A, 235, 131

 \bibitem[\protect\citeauthoryear{Homan et al. }{2002}]{Hom02} 
Homan J., van der Klis M., Jonker P.G., Wijnands R., Kuulkers E., Mendez  M., Lewin W.H.G., 2002, ApJ 568, 878

\bibitem[\protect\citeauthoryear{Homan et al.}{2007}]{Hom07}  
Homan J., et al., 2007, ApJ, 656, 420

\bibitem[\protect\citeauthoryear{Jahoda et al. }{1996}]{Jah96}
 Jahoda K, Swank J.H., Giles A.B., Stark M.J., Strohmayer T., Zhang  W., Morgan E., 1996, SPIE, 2808, 59

\bibitem[\protect\citeauthoryear{Koyama et al.}{2007}]{Koy07}
 Koyama K., et al., 2007, PASJ, 59, 23


\bibitem[\protect\citeauthoryear{ Kuulkers \& van der Klis}{1996}]{Kul96} 
Kuulkers E., van der Klis M., 1996, A\&A, 314, 567
\bibitem[\protect\citeauthoryear{ Kuulkers et al.}{1997}]{Kul97}
Kuulkers E., van der Klis M., Oosterbroek T., van Paradijs J., Lewin W.H.G., 1997, 287, 495

\bibitem[\protect\citeauthoryear{Lavagetto et al. }{2008}]{Lav08}
 Lavagetto G., Iaria R., D'Al A., Di Salvo T., Robba N.R., 2008, A\&A,  478, 181

\bibitem[\protect\citeauthoryear{Lin, Remillard \& Homan }{2009}]{lin09}
Lin D., Remillard R.A., Homan J., astroph/0901.0031


\bibitem[\protect\citeauthoryear{McClintock and Remillard}{2003}]{mc03}
McClintock J.E. \& Remillard R.A., 2003, to appear in {\it Compact Stellar X-ray Sources}, eds. W.H.G Lewin and M. van der klis, Cambridge University Press

\bibitem[\protect\citeauthoryear{Market \& Clark }{1975}]{Mar75}
Market T.H., \& Clark G.W., 1975, ApJ, 196, L55

\bibitem[\protect\citeauthoryear{McGowan et al. }{2003}]{Mcg03}
 McGowan K. E., Charles P.A., O'Donoghue D., Smale A.P., 2003, MNRAS,345, 1039

\bibitem[\protect\citeauthoryear{Mitsuda et al.}{1984}]{Mit84}
Mitsuda K., et al., 1984, PASJ, 36, 741
\bibitem[\protect\citeauthoryear{Mitsuda et al.}{1989}]{Mit89}
Mitsuda K., Inoue H., Nakamura N., Tanaka Y., 1989, PASJ, 41, 97
 \bibitem[\protect\citeauthoryear{Mitsuda et al.}{2007}]{Mit07}
 Mitsuda K., et al., 2007, PASJ, 59, 1

\bibitem[\protect\citeauthoryear{ O'Neill et al. }{2002}]{One02}
O'Neill P.M.,Kuulkers E., Sood R. K., van der Klis M., 2002, MNRAS, 336, 217

 \bibitem[\protect\citeauthoryear{Pakull }{1978}]{Pak78} 
Pakull M.W., 1978, IAU Circ., 3313

 \bibitem[\protect\citeauthoryear{Popham \& Sunyaev }{2001}]{Pop01}
 Popham R., Sunyaev R., 2001, ApJ, 547, 355

\bibitem[\protect\citeauthoryear{Psaltis Lamb \&  Miller}{1995}]{Psa95} 
Psaltis D., Lamb F.K., Miller G.S., 1995, ApJ, 454, L137
 \bibitem[\protect\citeauthoryear{Psaltis \& Lamb}{1997}]{Psa97}
 Psaltis D., Lamb F.K., 1997, ApJ, 488, 881

\bibitem[\protect\citeauthoryear{Smale, Homan \& Kuulkers}{2003}]{Sma03} 
Smale A.P., Homan J., Kuulkers E., 2003,  ApJ, 590, 1035
 \bibitem[\protect\citeauthoryear{Smale and Kuulkers}{2000}]{Sma00}
 Smale A.P., Kuulkers E., 2000, ApJ, 528, 702


 \bibitem[\protect\citeauthoryear{Takahashi et al.}{2007}]{Tak07}
 Takahashi T., et al., 2007, PASJ, 59, 35

 \bibitem[\protect\citeauthoryear{Titarchuk }{1994}]{Tit94}
Titarchuk L., 1994, ApJ, 434, 313
 
\bibitem[\protect\citeauthoryear{ van der Klis}{2000}]{Van00} 
van der Klis M., 2000, ARA\&A, 38, 717

\bibitem[\protect\citeauthoryear{Vrtilek et al.}{1990}]{Vrt90} 
Vrtilek S.D., Raymond J.C., Garcia M.R., Verbunt F., Hsinger G., Kurster M., 1990, A\&A, 235, 162

 \bibitem[\protect\citeauthoryear{White et al. }{1986}]{Whi86} 
White N. E., et al., 1986, MNRAS, 218, 129
 
 

 \bibitem[\protect\citeauthoryear{Wijnands et al. }{1997}]{Wij97}
 Wijnands R., van der Klis M., Kuulkers E., Asai K., Hasinger G.,  1997, A\&A, 323, 399







\label{lastpage}

\end{thebibliography}
\end{document}